\documentclass[twocolumn]{article}

\usepackage{geometry}
\usepackage{mathpazo}
\usepackage{amssymb}
\usepackage[fleqn]{amsmath}
\usepackage{graphicx}
\usepackage{epstopdf}
\usepackage[titletoc,title]{appendix}
\usepackage{gensymb}
\usepackage{bm}
\usepackage{authblk}
\usepackage{xcolor}

\usepackage{soul}

\renewenvironment{abstract}{%
  \small
  \begin{center}%
    {\bfseries \abstractname\vspace{-0.5em}\vspace{0pt}}%
  \end{center}%
  \quotation}
  {\endquotation}

\author[1,*]{Yeonsig Nam}
\author[2,3]{Majed Chergui}
\author[1]{Linda Young}
\author[1,*]{J\'er\'emy R. Rouxel}

\affil[1]{Chemical Sciences and Engineering Division, Argonne National Laboratory, Lemont, Illinois 60439, United States}
\affil[2]{Elettra-Sincrotrone Trieste, SS 14 - km 163.5, 34149 Basovizza, Trieste, Italy}
\affil[3]{Lausanne Centre for Ultrafast Science (LACUS), Ecole Polytechnique Fédérale de Lausanne (EPFL), ISIC, 1015 Lausanne, Switzerland}

\affil[*]{Corresponding authors: nam@anl.gov, jrouxel@anl.gov}

\title{Linear and nonlinear X-ray spectra of chiral molecules: X-ray Circular Dichroism, Sum- and Difference-Frequency Generation of fenchone and cysteine}

\date{\today}

\begin{document}

\twocolumn[
\maketitle 
\begin{abstract}     
Recent advancements in X-ray light sources at synchrotrons and X-ray free-electron lasers (XFELs) present exciting opportunities to probe molecular chirality using novel nonlinear spectroscopies with element-sensitivity. 
Circular dichroism (CD) and sum- and difference-frequency generation (SFG/DFG) are established techniques for probing molecular asymmetry in optical regime, with SFG/DFG offering unique advantages due to their background-free nature and independence from circularly polarized light sources. 
Extending them into the X-ray domain provides deeper insights into local structural asymmetries by leveraging the localized nature of core orbitals. 
In this work, we simulate X-ray absorption spectroscopy (XAS), X-ray circular dichroism (XCD) and nonlinear optical/X-ray SFG and DFG (OX SFG/DFG) signals for two prototypical chiral molecules, fenchone and cysteine. 
Our multi-reference simulations reproduce experimental data when available and reveal how novel X-ray spectroscopies exploit the site- and element-sensitivity of X-rays to uncover molecular asymmetry. 
The XCD spectra show strong asymmetries at chiral centers, while distant atoms contribute less. 
The OX SFG/DFG signals, under fixed resonant optical excitation, strongly depend on core transition and valence excitation energies. 
This dependence allows us to introduce two-dimensional (2D) chirality-sensitive valence-core spectroscopy, which provides insight into the overlap between valence orbitals and local molecular asymmetry. 
Finally, our estimates using realistic laser and X-ray pulse parameters demonstrate that such nonlinear experiments are feasible at XFELs, offering a promising tool for investigating the geometric and electronic structures of chiral molecules.

    \end{abstract}
\vspace{\baselineskip}
]

\section{Introduction}

Since its discovery nearly two centuries ago  \cite{vantomme2021pasteur}, molecular chirality has been characterized by optical spectroscopic techniques, which include electronic circular dichroism (ECD)  \cite{berova2000circular}, vibrational circular dichroism (VCD)  \cite{nafie1976vibrational}, optical rotatory dispersion (ORD)  \cite{crawford2007current, polavarapu2002optical} and Raman optical activity (ROA)  \cite{barron2004raman}.
They have provided ever-increasing sensitivity to discriminate enantiomers, molecules that are mirror images of one another but that cannot be superimposed, with asymmetry ratios as low as $10^{-6}$  \cite{berova2007application} and down to nanomolar concentrations  \cite{prabodh2021fluorescence}.
Efforts have been deployed to increase the signal strength of chiral spectroscopies, and new schemes such as photoelectron circular dichroism (PECD)  \cite{janssen2014detecting, turchini2017conformational} have been introduced using extreme ultraviolet (EUV) sources of radiation such as synchrotrons, XFELs or high-harmonic generation (HHG) where asymmetry ratios routinely approach 10\%. 
These novel probes have been used to monitor photoinduced chirality changes, both electronic and nuclear, in the time-domain \cite{baykusheva2019real, facciala2023time, wanie2024capturing}.
PECD has almost exclusively been applied to gas phase systems, although recent work has demonstrated it in the case of a chiral liquid of fenchone \cite{pohl2022photoelectron, malerz2022setup}.
Furthermore, in most cases, the incoming pulses used in these techniques interact with the valence electrons, which are delocalized over the molecule. 
Structural information on the local asymmetry within the molecule is thus only indirectly available.

\begin{figure}[h]
    \centering
    \includegraphics[width=0.5\textwidth]{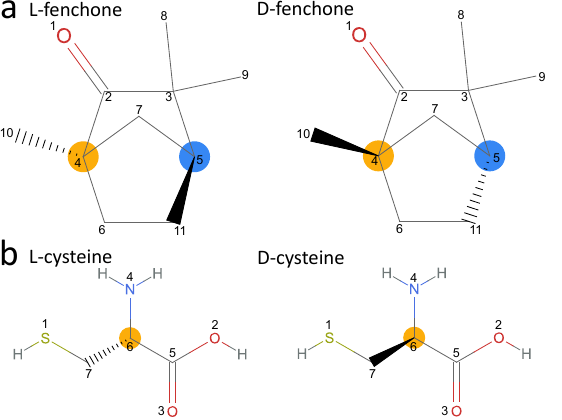}
    \caption{Skeletal formula of L- and D-fenchone ((1R,4S)-- and (1S,4R)--fenchone respectively) (a) and of L- and D-cysteine (b). The atom numbering is used throughout the text for peak assignment.
    The chiral centers are indicated by orange and blue circles.
    %% label numbers too small, need to reexport
    }
    \label{fig:fig1}
\end{figure}

Going into the X-ray domain offers additional structural insights into the description of chiral systems: a) X-rays resonant with core-transitions interact with localized core-electrons and thus probe selected atoms in the molecule. 
The local chirality associated with the element- and site-sensitivity of X-rays has been theoretically explored \cite{carra2000x, jiemchooroj2007near, kimberg2007calculation, jiemchooroj2008x, villaume2009circular, takahashi2015theoretical, oreshko2018calculation, freixas2023jacs,nam2022jacs}. 
The chiral response of atoms is found to depend on their chemical environment, thus adding a parameter to distinguish iso-elemental 
%\blue{how about "iso-elemental"? identical sounds confusing, same for the conclusions: MC, fine with me} 
but inequivalent atoms in a polyatomic molecule.

Implementing X-ray CD (XCD) experimentally started in the late 1990s when it became possible to generate circularly polarised light (CPL) at synchrotrons. 
The first studies were carried out using hard X-rays on oriented molecular crystals \cite{alagna1998x, goulon1998x, stewart1999circular, peacock2001natural}.
%\textit{(Give the refs: 10.1103/PhysRevLett.80.4799; https://doi.org/10.1021/jp001946y; https://pubs.acs.org/doi/10.1021/ja9920720; https://aip.scitation.org/doi/abs/10.1063/1.476046)} 
For ordered systems, it is the electric dipole / electric quadrupole interaction ($\bm \mu \bm q$, noted E1/E2) that dominates the CD signal \cite{peacock2001natural}. 
In the soft X-ray range, which is more interesting for the study of organic molecules as it covers the range of core transitions of light atoms (C, N, O, F, S, Cl, etc.), the challenge was the necessity to work under vacuum.
Therefore, most studies were carried out on films or powders of (bio)organic molecules \cite{nakagawa2005natural, tanaka2005first, izumi2013characteristic}. 
For isotropic ensembles, the electric quadrupole contribution vanishes upon orientational averaging and one needs to rely on electric dipole / magnetic dipole interactions ($\bm \mu \bm m$, noted E1/M1), making the signal more challenging to acquire because magnetic interactions are intrinsically weaker.
To date, only one work has recorded XCD signals on an isotropic, gas phase sample at the C K-edge \cite{turchini2004core} where the sensitivity to different carbons in the molecular structure was shown.

New capabilities at XFELs sources are providing opportunities for polarization studies on liquids.
Most XFEL experiments have so far focused on linear X-ray spectroscopy or diffraction, and proof-of-principle non-linear X-ray experiments have mostly been carried out on gas or solid phase samples \cite{chergui2023progress}, while extension to the liquid phase is still pending. 
With the advent of new schemes for the delivery of thin liquid solutions into vacuum \cite{ekimova2015liquid, kleine2018soft, fondell2017time, de2024sub}, 
soft X-ray absorption spectroscopy of liquid samples is now possible and and their study by soft X-ray CD studies can therefore be envisioned. 
% Because most investigated edges so far are K-edges, the dichroic contribution is expected to represent less than a percent of the total X-ray absorption signal.
Progress in polarization control of light \cite{poldi2020versatile, strempfer2022possibilities} and new light sources, such XFELs and HHG sources, open a more accessible path to novel chiral spectroscopies \cite{ordonez2018generalized, ayuso2019synthetic, rouxel2022molecular}. 
%In particular, the advent of free electron lasers (FELs) with their high-brilliance are enabling higher-order non-linear X-ray spectroscopies.
% Some of these non-linear methods are promising for the investigation of chirality.
In particular, second-order techniques \cite{glover2012x, tamasaku2014x, shwartz2014x, schori2017parametric} are particularly appealing in this respect.

\begin{figure}[h]
    \centering
    \includegraphics[width=0.48\textwidth]{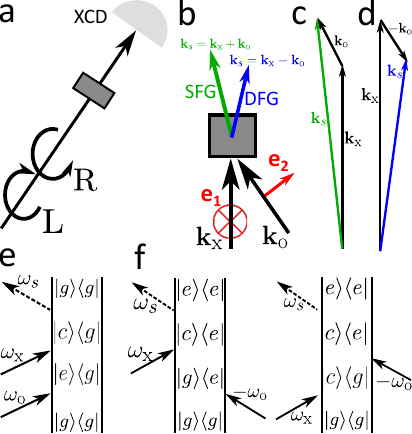}
    \caption{Geometry of a) X-ray Circular Dicrhoism (XCD) and; b) Optical/X-ray SFG/DFG detection schemes. c) and d) $\bm k$-vector diagrams for SFG and DFG, respectively. The wave vectors are not displayed in real scale, since the X-ray wave vector amplitude is much larger that the optical one. e) and f) Ladder diagrams contributing to the SFG and DFG spectra respectively.}
    \label{fig:fig2}
\end{figure}

It is well-known that $\chi^{(2)}$ techniques vanish in centrosymmetric media and, for that reason, they have successfully been used to probe surfaces and interfaces, but they are also a sensitive probe of asymmetry of the bulk of chiral media \cite{fischer2005nonlinear, lee2018femtosecond, rouxel2024manipulating}.
Moreover, non-collinear $\chi^{(2)}$ signals do not have an achiral background from lower multipoles and do not require incoming circularly polarized light (CPL).
This offers an alternative to CD, which suffers from weak signals overshadowed by a stronger absorption background.
Sum-frequency generation (SFG) and difference-frequency generation (DFG) measurements of chiral solutions have been demonstrated in the optical domain  \cite{rentzepis1966coherent,fischer2000three,belkin2001sum} but their widespread use has been limited, possibly due to the success of vibrational and electronic CD for enantiomeric discrimination. % due to the success of alternatives spectroscopic methods.
Here, we focus on sum- and difference-frequency mixing of an X-ray and an optical pulse, an effect which has already been demonstrated on diamond \cite{glover2012x}, but not on chiral liquids.
Such a mixing implies a large difference in the amplitude of their wave-vectors.
As a consequence, the Optical/X-ray (OX) SFG/DFG signals are emitted in almost the same direction as the incident X-ray beam, see the $\bm k$-diagram in Fig. \ref{fig:fig2}c, which calls for a spectral separation of the beams.

%{\color{red} Note that the incident angles $\theta_\text{o,SFG/DFG}$ and the emission angles $\theta_\text{s,SFG/DFG}$ are constrained by the phase matching conditions (see supplementary materials) that lead to:
%\begin{eqnarray}
%\theta_\text{o,SFG} &=& \arccos\Big[\frac{\lambda_\text{o}}{2n_\text{o}\lambda_\text{x}}\Big(\frac{\lambda_\text{x}^2}{\lambda_s^2}-1\Big)- \frac{n_\text{o}\lambda_\text{x}}{2\lambda_\text{o}}\Big] \label{eq:angle_SFG}\\
%\theta_\text{o,DFG} &=& \arccos\Big[\frac{\lambda_\text{o}}{2n_\text{o}\lambda_\text{x}}\Big(1-\frac{\lambda_\text{x}^2}{\lambda_s^2}\Big)+ \frac{n_\text{o}\lambda_\text{x}}{2\lambda_\text{o}}\Big]  \label{eq:angle_DFG}
%\end{eqnarray}
%For example, for a 5.6 eV + 277 eV SFG mixing process and assuming $n_\text{o}=1.45$, the phase matching condition imposes $\theta_\text{o}=47^\circ$ and $\theta_\text{s}=0.83^\circ$. %Similarly, a DFG process mixing 5.6 eV and 280.5 eV is phase matched with 
%$\theta_\text{o}=45.8^\circ$ and $\theta_\text{s}=-0.83^\circ$.}

While XCD is still a nascent methodology for molecular chirality, both in the gas and the liquid phase, a non-linear technique such as optical/X-ray SFG/DFG could be an exciting and complementary tool to probe local molecular chirality in solutions.
OX SFG/DFG come with advantages and inconveniences with respect to XCD.
First, they are electric-dipole only signals and thus, OX SFG/DFG do not rely on weaker higher order multipoles \cite{ayuso2022strong, rouxel2024manipulating}.
Additionally, there is no achiral contribution to the signal at the leading multipolar order and thus all emitted photons are informative on molecular chirality, unlike for CD measurement in which the achiral background constitutes $\sim$ 99\% of the detected signal.
The non-linear nature of the process gives additional control knobs to investigate molecular degrees of freedom. 
Here, we will consider the simplest case of coincident optical and X-ray wave mixing, but using a delay between the two pulses would enable further insights on the induced molecular dynamics.
% The additional insights from resonant chirality-sensitive X-ray spectroscopies allow the characterization of which atoms within a chiral molecule contribute most to the asymmetric response.

%In this work, we present numerical simulations of the XCD and XS/DFG signals of two chiral molecules in the gas phase, fenchone and cysteine, at two different absorption edges, namely, the carbon and oxygen K-edges.
In this work, we present numerical simulations of two  chiral-sensitive linear (XCD) and non-linear (OX SFG/DFG) signals for two chiral molecules (fenchone and cysteine), at their carbon and oxygen K-edges.
These two molecules are appealing for future experimental demonstrations in the liquid phase since they can be obtained as a pure enantiomer liquid (fenchone) or with relatively high concentration of enantiomers in water (cysteine, up to 28 g/100 mL at 25 °C. \cite{kleemann1985amino})
Fenchone is a naturally occurring compound used in the fragrance industry  \cite{strub2014stereochemistry} and it has regularly been studied as a prototype for chirality-sensitive spectroscopies. \cite{lux2012circular, ozga2016x, malerz2022setup, pohl2022photoelectron, facciala2023time}
The cysteine amino-acid is also an archetypical chiral molecule, which forms disulfide bonds yielding the cystine amino-acid \cite{coleman1968optical}, an important constituent of proteins.
Fenchone has two stereogenic centers, while cysteine has only a single chiral carbon, see Fig. \ref{fig:fig1}. 

First, we summarize signal expressions for both XCD and SFG/DFG, section \ref{sec:expressions}, and describe our ab initio quantum chemistry methodology, section \ref{sec:abinitio}.
Next, we present the XCD and nonlinear spectra, sections \ref{sec:xcd} and \ref{sec:chi2} respectively, and demonstrate how X-ray chiral spectroscopies give additional insight into the asymmetry distribution within the molecules.

\section{Methods}

\subsection{Signal expressions}
\label{sec:expressions}

XCD signals depend on the rotatory strength  \cite{berova2000circular} $R_{cg}=\text{Im}(\bm \mu_{gc}\cdot\bm m_{cg})$ where $g$ and $c$ are the electronic ground and core-excited states, respectively. 
XCD can be calculated using the multipolar field-matter interaction Hamiltonian truncated at the magnetic dipole order  \cite{craig1998molecular}:
\begin{equation}
H_\text{int}(t) = -\bm\mu\cdot \bm E(t) -\bm m\cdot \bm B(t)
\end{equation}
\noindent where $\bm \mu$ and $\bm m$ are the electric and magnetic transition dipoles respectively and $\bm E$ and $\bm B$ are the incoming electric and magnetic field amplitudes. 
The E1/E2 contributions in the XCD signals of disordered systems vanish upon rotational averaging \cite{rouxel2022molecular} since the quadrupolar tensor is symmetric.

Assuming an incident monochromatic beam, the XCD signal is given by (see SI, section 1):
\begin{multline}
S_\text{CD}(\omega) = A_L(\omega)-A_R(\omega)\\
= \frac{2}{3\hbar^2}N |E_0|^2\frac{4}{c}
\sum_c R_{cg} \frac{\Gamma_{cg}}{(\omega-\omega_{cg})^2+\Gamma_{cg}^2} 
\label{CDdef}
\end{multline}
\noindent where $A_{L/R}$ is the absorption signal of the left and right CPL, $\omega$ and $E_0$ are the incoming field frequency and amplitudes, $\omega_{cg}$ and $\Gamma_{cg}$ and the transition frequencies and lineshape broadening, respectively.
The signal is proportional to the number of molecules $N$ in the interaction region.
We further define the dissymmetry ratio (also named the Kuhn factor) $g=(A_L-A_R) / ((A_L+A_R)/2)$ that is commonly used to display CD spectra \cite{berova2011comprehensive}.

The use of a finite basis set in quantum chemistry methods causes the rotatory strength $R = \text{Im}(\bm \mu_{gc}\cdot \bm m_{cg})$ to be dependent on the choice of the coordinate system, when using length gauge.
Therefore, we use the velocity gauge to calculate to recover origin-invariant rotatory strengths:
\begin{eqnarray}
R_{cg} &=& -\frac{e^2\hbar}{2m} \langle g| \bm r |c\rangle \cdot \langle c| \bm r \times \nabla |g\rangle \\
&=& - \frac{e^2 \hbar^2}{2m^2} \frac{1}{\omega_{cg}} \langle g| \nabla |c\rangle \cdot \langle c| \bm r \times \nabla |g\rangle
\label{eq:vgR}
\end{eqnarray}
\noindent where $\bm \mu^\text{LG}_{ij} = -e \bm r_{ij}$, $\bm \mu^\text{VG}_{ij} = -i(e\hbar/m) (\bm p_{ij}/\hbar\omega_{ji})$, $\bm m = -(e/2mc) \bm r \times \bm p$.
The $1/c$ speed of light prefactor in the magnetic dipole definition has been included in Eq. \ref{CDdef} and highlights the weaker amplitude of CD signals with respect to the absorption background.
In the following, rotatory strengths are calculated using its velocity gauge formulation, Eq. \ref{eq:vgR}.

\begin{figure*}[t!]
    \centering
    \includegraphics[width=1\textwidth]{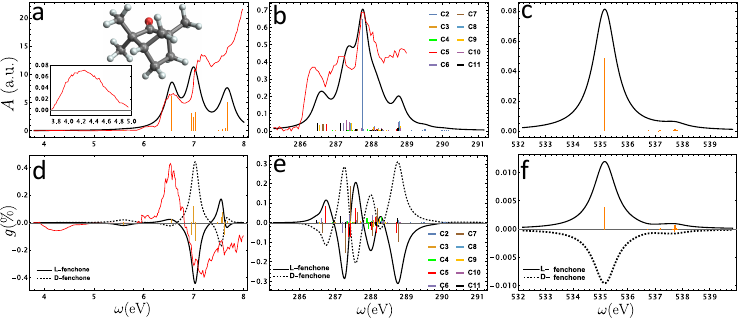}
    \caption{Absorption spectra (top) and CD dissymmetry spectra g(\%) (bottom) of fenchone. Spectra are computed in the UV (left), at the carbon (middle) and at the oxygen  (right) K-edges. Experimental absorption and CD spectra are shown as red traces for comparison \cite{pulm1997theoretical, ozga2016x, Singh2020fenchone}, with a zoom onto the spectral feature between 3.8 and 5 eV in inset.
     }
    \label{fig:fig3}
\end{figure*}

X-ray SFG signals do not vanish in the electric dipole approximation and thus we consider only the electric dipole transitions.
The homodyne-detected SFG signal, emitted with the wavevector $\bm k_s = \bm k_\text{x} + \bm k_\text{o}$, is given by (see Fig. \ref{fig:fig2}b and the SI, section 3):
\begin{multline}
S_\text{SFG}(\omega_x,\omega_o) = \frac{2N^2}{6^2\hbar^6}
\mathcal E_s^2 \mathcal E_\text{x}^2 \mathcal E_\text{o}^2
|\bm e_s^*\cdot(\bm e_x\times \bm e_o)|^2\\
\times\Big|
\sum_{ec} \bm \mu_{gc}\cdot(\bm \mu_{ce}\times\bm \mu_{eg})
f_{ceg}(\omega_s,\omega_\text{x},\omega_\text{o})
\Big|^2
\label{eq:XSFGdefmain}
\end{multline}
\noindent where $\mathcal E_s$, $\mathcal E_\text{x}$ and $\mathcal E_\text{o}$ are the amplitudes of the scattered, incident X-ray and incident optical beams respectively. $\bm e_s$, $\bm e_\text{x}$ and $\bm e_\text{o}$  are their corresponding polarization vectors. The indices $g, e$ and $c$ indicate the ground, valence-excited and core-excited states.
$f_{ceg}$ is a lineshape function defined in the SI.
Since nonlinear spectroscopies typically require high field intensities, OX SFG/DFG spectroscopies would typically be conducted with ultrashort pulses and thus the incoming fields $E_\text{o}$ and $E_\text{x}$ are not assumed to be monochromatic.
%{\color{red} (comment by Majed: but Eo can easily be made monochromatic)}.
% JR: correct, I rephrased the sentence accordingly
$\omega_\text{x}$ and $\omega_\text{o}$ are the central frequencies of the Gaussian envelopes for the X-ray and optical pulses, respectively, and $\omega_s$ is the detected frequency.
Similarly, the X-ray DFG signal (SI, section 4), emitted in the direction $\bm k_s = \bm k_\text{x} - \bm k_\text{o}$, is given by
\begin{multline}
S_\text{DFG}(\omega_\text{x},\omega_\text{o}) = \frac{2N^2}{6^2\hbar^6}
\mathcal E_s^2 \mathcal E_\text{x}^2 \mathcal E_\text{o}^2
|\bm e_s^*\cdot(\bm e_\text{x}\times \bm e_\text{o}^*)|^2\\
\times\Big|
\sum_{ec}
\bm \mu_{ec}\cdot(\bm \mu_{cg}\times \bm \mu_{eg}) 
g_{ceg}(\omega_s, \omega_\text{x}, \omega_\text{o})
\Big|^2
\label{eq:XDFGdef}
\end{multline}
Estimates of the photons flux with realistic XFEL parameters are given in the SI, section 5.

To maximize the signal, the polarization triple products $|\bm e_s^*\cdot(\bm e_x\times \bm e_o)|^2$ and $|\bm e_s^*\cdot(\bm e_x\times \bm e_o^*)|^2$  must be maximized.
This imposes the additional geometric constraint that the three polarizations involved in the three-wave mixing must be non-coplanar.
Momentum conservation $\bm k_x \pm k_\text{o} = \bm k_\text{s}$ constrains the incident angles $\theta_\text{o}$ and the emission angles $\theta_\text{s}$.
For example, for a 5.6 eV + 277 eV SFG mixing process and assuming $n_\text{o}=1.45$, the angles are given by $\theta_\text{o}=47^\circ$ and $\theta_\text{SFG}=0.83^\circ$. 
Similarly, a DFG process mixing 5.6 eV and 280.5 eV results in the following angles $\theta_\text{o}=45.8^\circ$ and $\theta_\text{DFG}=-0.83^\circ$.
Assuming a vertical linear polarization for the X-ray pump and a linear polarization in the horizontal plane for the optical pump, we get $|\bm e_s^*\cdot(\bm e_x\times \bm e_o)|^2 = 0.52$ for SFG at $\theta_\text{o}=47^\circ$ and $|\bm e_s^*\cdot(\bm e_x\times \bm e_o^*)|^2 = 0.53$ for DFG at $\theta_\text{o}=45.8^\circ$.

\subsection{Ab initio computations}
\label{sec:abinitio}

All ab initio computations were carried out using the MOLPRO package   \cite{werner2012molpro}. 
Equilibrium geometries of L/D-fenchone and L/D-cysteine were optimized at the MP2 level with the cc-pvdz basis set.

For fenchone, the valence excitations were calculated using the complete active space (multiconfigurational) self-consistent field (CASSCF) approach at the aug-cc-pVDZ/SA11-CASSCF(6e/15o) level of theory with the state averaging of eleven valence-excited states. 
The active space spanned from HOMO-2 to LUMO+11, including Rydberg orbitals, which were crucial for achieving accurate vertical excitation energies in close agreement with both TDDFT \cite{Singh2020fenchone} and CCSD \cite{10.1063/1.4973456} calculations.

Ten electronic core excitations at the oxygen and carbon K-edges were computed using restricted active space core-excited states calculations (RASSCF) for each of the carbon and oxygen 1s core molecular orbital at the aug-cc-pVDZ/SA10-RASSCF(8e/16o) by rotating the targeted 1s core orbital into the active space and restricting it to single occupation.
The second-order Douglas–Kroll–Hess Hamiltonian was used to account for relativistic corrections \cite{reiher2012relativistic}.

\begin{figure*}[t!]
    \centering
    \includegraphics[width=0.95\textwidth]{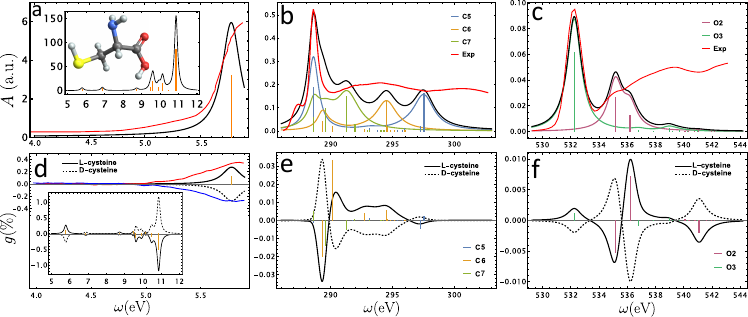}
    \caption{Absorption (top) and and CD dissymmetry spectra g(\%) (bottom) of cysteine. Spectra are computed in the UV (left), at the carbon K edge (middle) and at the oxygen K edge (right). Experimental absorption spectra are shown in red for comparison \cite{zubavichus2005innershell, wang2018surface}. For CD spectra, only the UV CD spectra have been recorded over a limited spectral range \cite{wang2018surface}, shown in red and blue in panel d for L- and D- cysteine respectively. The full spectral range computed is displayed in inset.}
    \label{fig:fig4}
\end{figure*}

A similar strategy was employed for the cysteine electronic structure calculations.
Ten valence excitations were computed at the cc-pVDZ/SA10-CASSCF(12e/10o), ranging from HOMO-5 to LUMO+3. Ten core excited states were computed at the cc-pVDZ/SA10-RASSCF(14e/11o).

%In both cases, carbon and oxygen atoms were numbered by order of increasing energy for the atomic-like molecular orbital energy levels. %\blue{original sentence: the first ground-to-core transition within their respective XAS spectra.}
The electric transition dipoles were computed using Eq. \ref{eq:vgR} with the matrix elements of the velocity operators that are computed by the VELO keyword in MOLPRO.
The origin-invariance was verified by randomly translating and rotating the molecular coordinates and checking that the computed rotatory strength was not appreciably modified.
%The computed fenchone (UV: Fig.\ref{fig:fig4}a, C K-edge: Fig.\ref{fig:fig4}b, O K-edge: Fig.\ref{fig:fig4}c) and cysteine (UV: Fig.\ref{fig:fig4}d, C K-edge: Fig.\ref{fig:fig4}e, O K-edge: Fig.\ref{fig:fig4}f) CD spectra were computed using the same phenomenological broadenings and energy shifts defined for the absorption spectra. %\blue{(do we need this last sentence??)}

Our choice of multi-reference electronic structure theory calculations and basis sets shows good agreement with both experimental data and other high-level simulations. 
For instance, our carbon K-edge spectra (Figure~\ref{fig:fig3}b), better aligns with the experimental spectrum \cite{ozga2016x} than the TDDFT calculations in ref. \cite{ozga2016x} using the PC3 basis set, which did not include augmented basis functions. 
The UV (Figure~\ref{fig:fig4}a) and O K-edge X-ray absorption spectra (Figure~\ref{fig:fig4}b) for cysteine are consistent with restricted energy window time-dependent density functional theory (REW-TDDFT) calculations\cite{YuZhang2012}. 
It is worth noting that vertical excitation energies, approximated as Kohn-Sham eigenvalues for HOMO-LUMO transitions, tend to be significantly underestimated \cite{Maul2007}.
Similarly, static exchange (STEX), a single Slater determinant theory, does not properly take into account electron correlation and the non-orthogonality of two excitation manifolds, which can result in spurious transition dipole moments between states. As a result, STEX is less suitable for obtaining experimentally comparable X-ray absorption spectra where electron correlation plays a crucial role.
A detailed comparison between REW-TDDFT and STEX for core-excited state calculations can be found in Ref. \cite{YuZhang2012}.
Given that our XAS spectra at both the C and O K-edges agree well with experiments, we are confident in the reliability of our XCD spectra for cysteine at the C K-edge (Figure~\ref{fig:fig4}e), despite minor discrepancies compared to the STEX calculations by Plashkevych et al.\cite{Plashkevych1998}.

\section{Absorption and circular dichroism spectra}
\label{sec:xcd}

\textbf{Fenchone spectra.}
The simulated UV absorption spectrum using a phenomenological broadening of 0.15 eV is compared in Fig. \ref{fig:fig3}a with the experimental spectrum from ref. \cite{pulm1997theoretical}. 
An energy shift of 0.7 eV was applied to all transitions energies to match the UV experimental spectra.
A good agreement with experimental spectra \cite{Singh2020fenchone, pulm1997theoretical} is observed for the features above 5.5 eV. 
However, the weak experimental feature between 4 and 5 eV reported by Pulm et al\cite{pulm1997theoretical}, and shown as an inset in Fig. \ref{fig:fig3}a, is not recovered in our calculation.
The origin of this discrepancy is not clear and may stem from the level of theory used. Note that Pulm et al. used a DFT level of theory with the B3LYP hybrid functional and also get significant differences with their experimental data.

Figs. \ref{fig:fig3}b and c show the C and O K-edge spectra with a phenomenological broadening of 0.2 and 0.5 eV. An energy shift of 2.2 was applied to the C K-edge spectra, none was applied to the O K-edge ones in the absence of experimental reference.
An additional edge-jump function is added for comparison in SI, Fig. S3.
The labeling of atoms refers to Fig. \ref{fig:fig1}. 
The strong X-ray absorption at the C2 K-edge (Fig. \ref{fig:fig3}b) is attributed to the significant orbital overlap between the C2 1s electron and the C2$=$O $\pi$ arrival orbital, resulting in a large oscillator strength. While optical/UV spectroscopies probe the global electronic structure of the molecules, the appeal of X-ray CD spectra is that they provide insight into the local asymmetry of molecules thanks to the localized nature of core orbitals. 

The computed and experimental UV CD spectra of fenchone are compared in Fig. \ref{fig:fig3}d. The simulated spectrum displays significant positive and negative features at 6.6 eV and 7 eV, respectively, but their relative amplitudes deviate from the experimental CD spectrum \cite{pulm1997theoretical}. 
Similarly, the weak negative feature calculated at 5.6 eV appears at 4-4.6 eV in the experimental spectrum \cite{pulm1997theoretical}.

Fig. \ref{fig:fig3}e and f show the calculated X-ray CD spectra of fenchone in the regions of the carbon K-edge (panel e) and oxygen K-edge (panel f). 
Note that experimental XCD spectra of liquid solutions are not available to date for either fenchone or cysteine.
Due to the presence of two chiral carbon atoms in fenchone and its polycyclic structure, the C K-edge CD spectrum (Fig. \ref{fig:fig3}e) is quite complex. It contains additional and valuable information compared to the X-ray absorption spectrum. 
% In Fig. \ref{fig:fig3}e, 
Indeed, the strongest XCD features appears at the C5 chiral center and the C7 bridge atom.
The  C2 atom, which is brightest in the XAS, contributes significantly less in the XCD. In Fig. S2, individual contributions to the C K-edge CD spectrum are displayed to highlight these observations.

The XCD spectrum at the oxygen K-edges displays fewer features than its carbon counterpart. This is due to the fact that there is a single oxygen atom in fenchone which thus gives insight only on the chirality in its local environment.
This is similar to previous work in which a single X-ray chromophore was used to probe the local chiral structure \cite{zhang2017x}.
The asymmetry ratio is also smaller (\~0.01 \%) compared to the carbon K-edge XCD, indicating a more symmetric environment at the vicinity of the oxygen atom.
\\

\noindent \textbf{Cysteine spectra.}
We now discuss the cysteine spectra displayed in Fig. \ref{fig:fig4}.
The simulated UV absorption spectrum was computed with a 0.1 eV phenomenological broadening without energy shift and is shown in Fig. \ref{fig:fig4}a. 
The agreement between calculated and experimental spectrum (available only below 6 eV \cite{wang2018surface}) is good.
% No experimental reference spectra are available for cysteine below 200 nm. 
% Their measurements in the liquid phase are complicated by the absorption of water. 
An additional broadening of the peaks is expected due to the presence of multiple conformers in the solution phase \cite{Maul2007}.
Turning now to the C and O K-edge spectra, Figs \ref{fig:fig4}b and \ref{fig:fig4}c, these were shifted by 2.7 and 3.2 eV and broadened by 0.2 and 0.5 eV, respectively, to match the experimental spectra \cite{zubavichus2005innershell,Maul2007}.
Our computation includes only contributions from bound states and does not include photoionized electrons contributing above the edge. The insight into the local asymmetry of the molecules is simpler to observe in Cysteine thanks to its single origin of the chiral center. For instance, in Fig. \ref{fig:fig4}e the strongest contribution to the C K-edge XCD signal originates from the C6 atom at the chiral center of cysteine. In general, CD signals decrease with increasing distance from the chiral center \cite{zhang2017x}. For example, the C5 atom shows the largest absorption at 288.6 eV, Fig. \ref{fig:fig4}b, while it is the C6 atom that dominates the XCD spectrum. %that is dominated by the C6 atom.
In the oxygen spectra, the O3 atom dominates the XAS, while O2 dominates the XCD. In this case too, the oxygen XCD displays a smaller asymmetry ratio than observed at the carbon K-edge.

\section{Optical/X-ray SFG and DFG spectra}
\label{sec:chi2}

\begin{figure}[h!]
    \centering
    \includegraphics[width=0.4\textwidth]{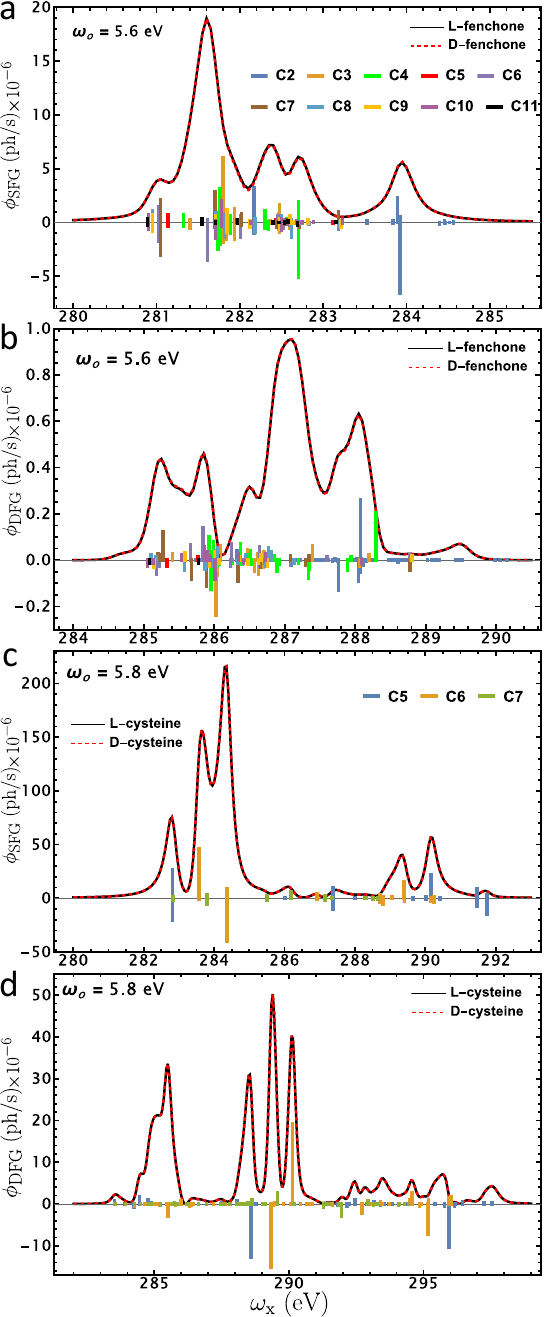}
    \caption{X-ray Sum-Frequency Generation (a and c) and Difference-Frequency Generation (b and d) on fenchone and cysteine at the carbon K-edge, see Eq.\ref{eq:XSFGdefmain} and Eq.\ref{eq:XDFGdef}.
    The UV pulse central frequency is indicated in inset. 
    %\blue{shouldn't we plot O K-edge here? or at least at the SI?}
    % JR: yes maybe, need to export it out
    }
    \label{fig:fig5}
\end{figure}

Fig. \ref{fig:fig5} shows the Optical/X-ray SFG/DFG spectra computed with $\omega_\text{x}$ scanned over the C K-edge. 
The UV pulses being mixed with the X-ray pulse have a photon energy $\omega_\text{o}$  matching a UV transition, 5.56 eV in fenchone and 5.8 eV in cysteine.
As explained in SI section S5, the calculated photon flux of the signals is $10^6 - 10^8$ photons/s, indicating that the detection of this signal is possible with current capabilities, even when taking into account reabsorption by the solvent and the efficiency of crystal analyzers.

Fig. \ref{fig:fig5} shows the C K-edge SFG and DFG spectra.
As shown in the SI (S3 and S4), these two signals involve inequivalent excitation pathways, see Fig.\ref{fig:fig2}, e and f. 
The SFG pathway presented here involves a valence excitation $g\rightarrow e$ then followed by a valence-to-core transition $e \rightarrow c$. 
Thus, the resonant process involves an X-ray photon energy below the main edge and red shifted by the chosen valence excitation energy (5.56 eV for fenchone and 5.8 eV for cysteine). 
This could be advantageous since the competing linear absorption is off-resonant for the X-ray pulse in this case.
The DFG pathway involves interactions both on the bra and ket of the observable expectation value (see Fig. S3 in SI) and two pathways with a different time ordering are possible by interacting first with the optical photon ($-\omega_o$) then with the X-ray photon ($\omega_X$), or the reverse.
Here, the X-ray photon energy is tuned to the lowest core absorption $g\rightarrow c$ so that the off-resonant contributions are neglected, but the signal emitted by the DFG process generates photons below the edge corresponding to the transition $c\rightarrow e$.
The advantage here would then be on the detection side, since one would expect emission in a spectral domain in which the molecule does not emit usually.
This is advantageous for a detection relying on the spectral dispersion the signal, especially since the emission direction $\bm k_s$ is expected to be close to the incident beam $\bm k_\text{x}$.

As observed with the XCD spectra, the OX-SFG/DFG spectra carry information about the carbons contributing to the chiral nonlinear response, but with the additional window of the valence excited states.
First the SFG/DFG signals are proportional to the triple product of transition dipoles (Eq. \ref{eq:XSFGdefmain} and \ref{eq:XDFGdef}).
Their expressions contain two resonant terms, leading to a situation where the X-ray photon may be resonant while the optical photon is not. Consequently, OX SFG/DFG signals can remain weak even when a given triple product is large.
To account for this, the sticks in Fig. \ref{fig:fig5} are weighted by the resonant propagator and are thus given by $\bm \mu_{cg}\cdot(\bm \mu_{ce}\times \bm \mu_{eg})/|\omega_\text{o} -\omega_{eg}+i\Gamma_{eg}|$ for SFG and by $\bm \mu_{ec}\cdot(\bm \mu_{cg}\times \bm \mu_{eg})/|\omega_\text{o} -\omega_{eg}+i\Gamma_{eg}|$ for DFG. 
The sticks themselves are positioned at the resonances of the other propagator when $\omega_\text{x}$ is varied, $\omega_{cg}-\omega_\text{o}$ for SFG and $\omega_{ce}+\omega_\text{o}$ for DFG.
The triple products can be positive or negative and are first summed together. 
The final signal is obtained by squaring the amplitude of this sum.
%\st{the final signal is an amplitude square of the sum of their contributions.}
Because of this, some contributions can interfere and a spectral feature in the broadened spectrum is not as directly connected to the stick spectrum like linear absorption and oscillator strengths are.
In Fig. \ref{fig:fig5}a, the SFG signal of fenchone display strong features originating from the chiral center C4 at 282.8 eV, from C3 at 281.8 eV and from C2 at 283.9 eV.
For DFG, the spectrum is also dominated by the carbons C2, C3 and C4 in different regions of the spectrum.
In cysteine, it can be see that the main contributions to the SFG and DFG signals are from the chiral carbon C6, while the atom C5 contributes also significantly in other parts of the spectrum. 
The carbon C7 does not contribute significantly to the signal.
% For example, the large triple product visible at 283.7 eV in Fig. \ref{fig:fig5}a correspond to a contribution that would be resonant with a different optical photon energy.

\section{Discussion}

The CD spectra in Fig. \ref{fig:fig3} and \ref{fig:fig4} are displayed in percent of the asymmetry ratio (100 CD / Abs)\cite{berova2000circular}.
The fenchone C K-edge CD, Fig. \ref{fig:fig3}e, has amplitudes reaching $\pm$0.3\%, a value not uncommon for CD spectra.
While asymmetry ratios as small as 0.01\% were recorded in the optical regime \cite{berova2007application}, the signal-to-noise level required to measure such low variations in the X-ray absorption spectra have long been out of reach for synchrotron-based XCD experiments and have long limited their possibilities \cite{turchini2004core}.
Recent progress in source quality and beam line design calls for renewed attempts at recording XCD spectra of gases and solutions \cite{poldi2020versatile, strempfer2022possibilities}.
%The O K-edge spectrum with asymmetry ratio in the range of 0.01\% may be out of reach of current capabilities. 
These challenges of XCD spectroscopy are made even more complex, if one were to envision time-resolved studies in a pump-probe fashion, as this would imply a double-differential measurement. This is where exploring non-linear X-ray spectroscopies may become attractive.
%This can be understood by observing that their distance from the chiral centers.{\color{red} what do you mean?}

\begin{figure}[h!]
    \centering
    \includegraphics[width=0.5\textwidth]{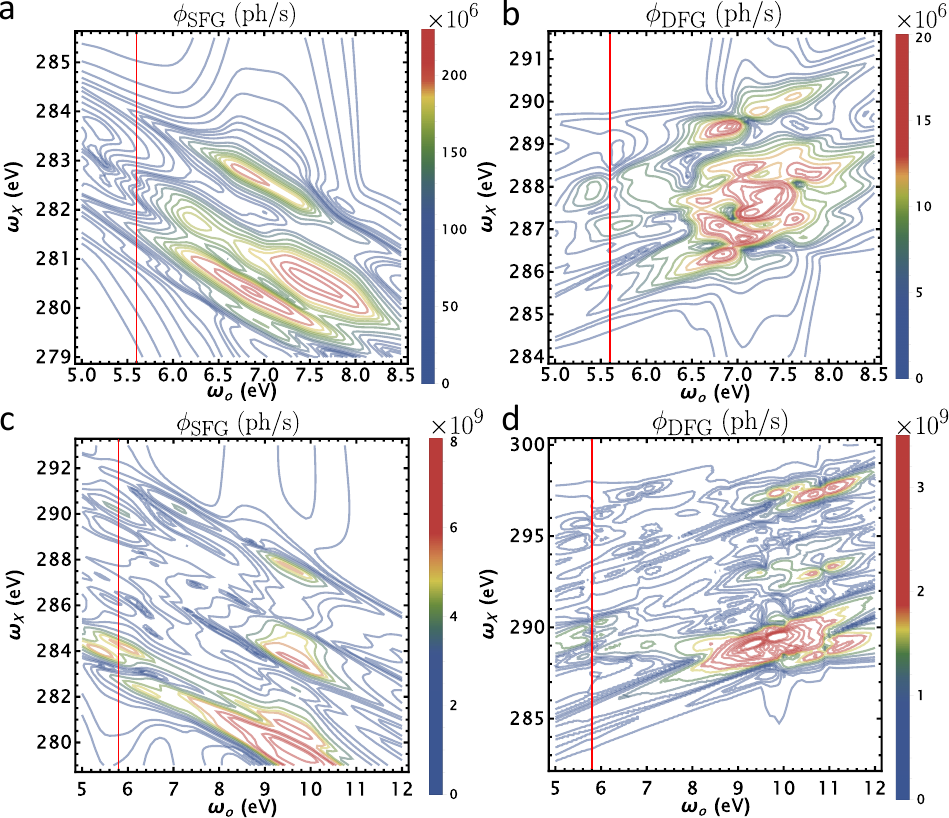}
    \caption{Valence-core two-dimensional (2D) optical/X-ray (OX) SFG and DFG spectra of fenchone (a and b) and cysteine (c and d). The vertical red lines indicate the slices displayed in Fig. \ref{fig:fig5}.
    }
    \label{fig:fig6}
\end{figure}

The full nonlinear response can be appreciated by scanning both the optical and the X-ray frequency.
This gives the opportunity to introduce the concept of valence-core two dimensional (2D) spectroscopy.
The 2D OX SFG and DFG spectra are shown in Fig. \ref{fig:fig6} in panels a and b for fenchone and c and d for cysteine.
The optical excitation energy is given in the x-axis, and it spans the region of different valence excitations. %, horizontal axes in Fig. \ref{fig:fig6}, 
The core excitation energies are shown on the y-axis and they span the different transitions around the C K-edge.
The SFG and DFG have an opposite slope due to the $1/(\omega_s-\omega_{cg}+i\Gamma_{cg})$ resonant factors where $\omega_s = \omega_\text{x} \pm \omega_\text{o}$. 
The spectra give insights into how much certain valence excitations overlap with the carbon core orbitals. 
This further depends on the asymmetry of the local structure in the vicinity of the carbon core orbital considered. 
If the molecular structure is known, the SFG/DFG spectra provide information on the location of the valence-excited electronic density over the molecule.
These 2D valence-core (VC) spectra represent a promising tool for getting information about the chiral molecular structure and its underlying electronic structure.

Signal fluxes for the 2D VC spectra are given in photon/second (ph/s) by taking into account realistic beam parameters for the optical laser and the XFEL pulses (optical: 20 $\mu$J, 25 fs, waist of 50 $\mu$m, $\sigma_\text{o} = 0.04$ eV. X-ray: 10 $\mu$J, 10 fs, waist of 20 $\mu$m , $\sigma_\text{x} = 0.25$ eV). 
We have considered a sample thickness of 2 $\mu$m, a repetition rate of 100Hz and a concentration of 2M for solvated cysteine.
For fenchone, a pure chiral liquid was considered.
The different signals are in the order of 1 to 100$\times 10^6$ ph/s, indicating possibilities to carry out the experiment at state-of-the-art XFEL facilities (the above parameters correspond to $2.2\times 10^{15}$ ph/s and $2.1\times 10^{13}$ ph/s for the optical and X-ray incident flux respectively).
For practical implementations, one would have to consider the efficiency of the spectrally-resolved detection.
Additionally, these signals decrease quickly off-resonance.
%The emitted photon flux is quickly decreased if off-resonant processes are considered. 
The recent development of high repetition rate FELs such as LCLS-II or EuXFEL could ease the detection of these nonlinear signals.

Finally, we address the enantiomer discrimination in OX SFG/DFG techniques.
The discussed homodyne detection scheme results in an amplitude squared expression for the signals and thus opposite enantiomers emit the same OX SFG/DFG signals, see Fig. \ref{fig:fig5}.
The homodyne-detected OX SFG/DFG is thus chirality-sensitive (non-zero only in chiral systems) but not enantioselective (opposite signal sign for opposite enantiomer).
Enantioselectivity can be achieved by adding a local oscillator beam in a heterodyne detection scheme.
In the latter, a phase-locked local oscillator is used to interfere with the signal, to recover its phase and to get access to the chiral pseudo-scalar involved in the SFG/DFG processes.
However, the low temporal coherence of intense X-ray pulses at XFELs makes heterodyne detection out of reach for now, while homodyne detection can readily be implemented.
% The simpler and readily available alternative is homodyne detection, in which the intensity of the signal is detected but its phase is lost.
% Homodyne-detected SFG/DFG is thus sensitive to molecular chirality, as it vanishes for achiral systems, but it cannot distinguish opposite enantiomers.
This limitation of homodyne detection is not a deterrent for the technique since other, well-established, complementary optical methods exist for the purpose of enantioselectivity. 

% In the homodyne-detection mode, signals are given as an amplitude-squared and the phase of the emission is thus lost, making it impossible to discriminate opposite enantiomers. 
% The enantiomeric discrimination could be recovered by carrying out a heterodyne-detection of the SFG photons, since the triple product $\bm \mu_{gc}\cdot(\bm \mu_{ce}\times\bm \mu_{eg})$ in Eq. \ref{eq:XSFGdefmain} is a pseudo-scalar with an opposite sign for each enantiomer.

\section{Conclusion}

In conclusion, we performed ab initio electronic structure calculations of CD and the Optical/X-ray SFG/DFG spectra for fenchone and cysteine at the carbon and oxygen K-edges.
XCD computations require the inclusion of magnetic transition dipole matrix elements, and rotatory strengths must be calculated using velocity gauge electric dipole matrix elements to ensure their origin invariance.
Chiral spectroscopies with resonant X-rays provide additional information relative to optical spectroscopies as transitions originate from atoms located in an asymmetric environment. 
We found the CD intensity strongest at the chiral center for both fenchone and cysteine, and it decreases with distance from it, in general.
In fenchone, which has two chiral centers, we observed that the two chiral carbons do not contribute equally to the XCD signal, in line with their different asymmetric environments. This highlights the ability of XCD as a tool to distinguish iso-elemental but inequivalent atoms in molecular systems.
XCD signals are small compared to the X-ray absorption background and have recently been reachable experimentally thanks to the increased stability and brilliance of circularly-polarized synchrotron-based X-ray sources.

Even-order nonlinear spectroscopies are alternatives probes. 
As a phase-matched non-collinear process, the OX SFG/DFG signal is emitted in a different direction than the transmitted exciting pulses and thus does not suffer from a large achiral background at the same photon energy as XCD does.
When a dispersive detection is added, the transmitted X-ray pump can be removed and the signal is truly background free.
It does not require circularly-polarized X-rays, but being a non-linear effect, the signal's absolute amplitude greatly profits from intense sources and thus XFELs are ideally suited for X-ray SFG and DFG experiments on chiral liquids.
Finally, all-X-ray $\chi^{(2)}$ techniques \cite{serrat2023resonant} represent the next step in the above described efforts aimed at probing molecular structure, and they rely on the availability of multipulse multicolor FEL schemes, which are becoming increasingly possible \cite{guo2024experimental}. 
% \textit{add: https://doi.org/10.1021/acs.jpca.1c06950; https://dx.doi.org/10.1088/1361-6455/ad0f39}

%{\color{red} refs 11 to 13 are cited as experimental work but they are theoretical papers, ref. 18 is XMCD which does not apply here. In figure 3 top, why do you show the fenchone spectrum starting at 5 eV, while there is an important band at 4 eV, which you also discuss.}.

%%%%%%%%%%%%%%%%%%%%%%%%%%%%%%%%%%%%%%%%%%%%%%%%%%%%%%%%%%%
%%%%%%%%%%%%%%%%%%%%%%%%%%%%%%%%%%%%%%%%%%%%%%%%%%%%%%%%%%%
\section*{Acknowledgements}
\addcontentsline{toc}{section}{Acknowledgements}

This work was supported by the U.S. Department of Energy (DOE), Office of Science, Basic Energy Science (BES), Chemical Sciences, Geosciences and Biosciences Division (CSGB) under Contract No. DE-FOA-0003176 for the study on linear signals (XAS, XCD) and Contract No. DE-AC02-06CH11357 for the study on nonlinear signals (SFG/DFG).
MC acknowledges support of the European Research Council via the Advanced Grant CHIRAX (ID 101095012).

\bibliographystyle{unsrt}
\bibliography{biblio}

%\printbibliography

\end{document}